# The effect of strain on the thawing of the hidden state and other transitions in *1T*-TaS$_2$.

Damjan Svetin, Igor Vaskivskyi, Petra Sutar, Jan Gospodaric, Tomaz Mertelj, Dragan Mihailovic
*Jozef Stefan Institute, Jamova 39, SI-1000 Ljubljana, Slovenia*

## Abstract

We investigate the effect of 2-dimensional (in-plane) strain on the critical transition temperature $T_H$ from the photoexcited hidden state in *1T*-TaS$_2$ thin films on different substrates. We also measure the effect of in-plane strain on the transition temperature $T_{c2}$ between the nearly commensurate charge-density wave state and the commensurate state near 200 K. In each case, the strain is caused by the differential contraction of the sample and the substrate, and ranges from 0.5 % compressive strain (CaF$_2$) to 2 % tensile strain (sapphire). Strain appears to have an opposite effect on the *H* state and the *NC-C* state transitions. $T_H$ shows a large and negative strain coefficient of $dT_H/de$ = - 8900±500 K, while $T_{c2}$ is not strongly affected by tensile strain and shows a positive coefficient for compressive strain, which is opposite to the effect observed for hydrostatic pressure.

## Introduction

Control of phase transitions by external perturbations can be potentially very important for device applications, particularly for making nonvolatile memory devices. The possibility of using macroscopically ordered states to hold information leads to inherently bistable systems if different states can be used which coexist under ambient conditions. The problem is how to reversibly switch states in such devices, while avoiding phase mixtures common in systems with first order transitions. A different possibility is offered by the recent discovery of switching to a hidden (*H*) state in *1T*-TaS$_2$, which is not otherwise accessible under thermodynamic conditions, but is possible under non-equilibrium conditions. In this material switching of a charge-ordered macroscopic quantum state is caused by a single ultrashort laser pulse, where the material switches from a non-conducting charge density wave Mott insulator to a conducting metallic state with different electronic order [1]. This material exhibits a number or competing ground states, with transitions from a metal to an incommensurate (IC) charge density wave state (at $T_{c0}$=550 K), then a transition to a nearly commensurate (NC) state (at $T_{c1}$=350 K), and finally to a commensurate (C) Mott-insulating state $T_{c2}$=180~220 K [2], but the hidden state is reachable only under non-equilibrium conditions.

The origin of the remarkable stability and uniformity of the switched *H* state is still not experimentally clarified in detail, and the detailed order of the hidden state is not yet known, so investigation of the system properties in response to external perturbations is paramount in understanding the underlying phenomena. The switching in this system was observed to be stable at temperatures below ~ 60 K, which is too low for widespread applications, so investigations into the detailed mechanism which stabilises the hidden state is of great importance, especially if it could be controlled externally.

It's known that the application of external pressure in high-temperature superconductors can lead to dramatic increases in the critical temperature [3]. In 1T-TaS$_2$, the application of hydrostatic pressure leads to systematic changes its functional properties, suppressing the transition from a nearly-commensurate CDW state to the commensurate (C) state, and eventually causing the material to become superconducting at temperatures below 5 K [4]. So it would be clearly interesting to investigate if the critical temperature associated with the transition from the hidden state to the visible ground state is influenced by external pressure. Apart from hydrostatic pressure, it has been shown that anisotropic stress can be caused by lattice constant mismatch between the thin quasi-epitaxial single crystalline films and the substrates on which they are grown, in which case both tensile and compressive stress can lead to changes in functional properties [5,6,7].

In this paper we present a study of the effect of tensile and compressive strain on the hidden state transition temperature as well as the transition from the nearly commensurate state to the commensurate Mott state in thin films of 1T-TaS$_2$ applied by different substrates. Anisotropic strain within the thin film sample is caused by a mismatch of the thermal expansion of thin films of 1T-TaS$_2$ and selected substrates on which they are deposited. We measure the electrical resistivity to detect the presence of the switching to and from the hidden state.

## Methods

The samples were deposited on sapphire (Al$_2$O$_3$), MgO, quartz (SiO$_2$), BK7 glass and CaF$_2$ substrates by exfoliating with Scotch tape and re-depositing the exfoliated flakes directly onto different substrates. The thicknesses of the films deposited in this investigation are between 50 and 100 nm. The crystal *c* axis of the samples is perpendicular to the plane of the substrate. Typical lateral dimensions were 100 µm, allowing four gold contacts to be deposited over the samples using laser direct photolithography using an LPKF Protolaser LDI laser writer as shown in Figure 1. The contacts were made with a Au/Pd intermediate layer before sputtering of 100±20 nm thick gold electrodes. The Van der Waals adhesion of the ultra-thin 1T-TaS$_2$ flakes to the substrates is assumed to be sufficiently good to avoid slippage upon cooling (which is evident from the smooth resistance curves during cooling). The substrate materials were chosen such that both compressive and tensile strain can be measured.

The crystal structure of 1T-TaS$_2$ in the absence of the small perturbation caused by the CDW is near-hexagonal (space group *P-3m1*, point group *D$_{3d}$*) in the relevant temperature range, while the substrates are either trigonal (sapphire and quartz), cubic (CaF$_2$ and MgO) or amorphous (BK7 glass). In the former two cases the stress is anisotropic in the plane of the thin film, while in the amorphous case the stress is assumed to be isotropic. The strain within the sample is related to the stress imposed by the substrate via the elasticity tensor, which in the case of hexagonal symmetry has 5 independent components [8]. The orientation of the sample axes with respect to the substrate crystal axes is assumed to be random with our method of deposition, so the applied stress may vary from sample to sample, depending on direction of the 1T-TaS$_2$ crystal axes with respect to the substrate crystal axes. To average out the effect of the in-plane anisotropy we have made measurements on a number of samples for each substrate material, thus averaging out the different stress tensor components arising from different sample orientation relative to the substrates.

The strain $e = \Delta L/L$ as a function of temperature for different substrates and bulk $1T$-TaS$_2$ material (in the $a$-$b$ plane) are shown in Figure 2. The relative strain of $1T$-TaS$_2$ with respect to the substrate $\Delta L_{1T\text{-}TaS2}/L - \Delta L_{substrate}/L$ is also plotted for different substrates (Fig. 2b)).

To switch to the $H$ state we expose the sample to a single ultrashort 35 fs laser pulse from an amplified Ti:Sapphire laser (Coherent Legend) with an energy 3 mJ/cm$^2$ at 30K (Stojchevska et al, 2014). The sample is exposed from the back through the substrate to ensure that the contact region is thoroughly illuminated, while four contacts are made on the top of the sample as shown in Figure 1 for the case of sapphire. We monitor the system properties by measuring the sample resistance upon warming through the transition to the C state.

# Experimental results

In Figure 3 we show the temperature dependence of the resistivity $R(T)$ of samples on representative substrates. CaF$_2$ provides the maximum compressive strain, while sapphire provides tensile strain. A shift of $T_{c2}$ is clearly evident. A spread of values is also evident, which - amongst other factors - comes from the random orientation of the sample with respect to the substrate.

In Figure 3 b) we show the transition temperatures $T_{c2}^c$ and $T_{c2}^w$ measured upon cooling and warming respectively averaged over 5 samples for each substrate. The strain is calculated at the actual transition temperature in each case, assuming isotropic stress. On cooling, we see that the effect of stress induced by the substrate on $T_{c2}^c$ is not significant for sapphire, MgO, polycrystalline SiO$_2$ and BK7 glass substrates, while for compressive strain (CaF$_2$), the effect is significant, where $T_{c2}^c$ is nearly 30 K higher. On warming the effects of the substrate changes $T_{c2}^w$ less than 10 K with tensile strain, with the exception of is CaF$_2$, which has a slightly higher $T_{c2}^w$. We conclude that the effect of stress induced by the substrate appears to be significant (20 ~ 30K) for compressive strain, and small for tensile strain.

In Figure 4 we show the resistivity through the $H$ state transition on warming as a function of strain. In each case, the sample is first cooled slowly from room temperature to 25 K, whereupon it was exposed to a 35 fs laser pulse to switch it to the $H$ state. The sample is then slowly heated while measuring $R(T)$. A systematic trend appears to be present, whereby $T_H$ decreases with increasing positive strain, such that $T_H$ is ~ 20 K lower on a CaF$_2$ substrate than on sapphire ($T_H$ = 57 K ± 5 K), and other substrates are in between.

# Discussion

Systematic slippage of the sample on the substrate can be probably excluded for the following reasons. Apart from the obvious absence of any discontinuities in the $R(T)$ curves, the systematic behaviour of $T_{c2}$ and $T_H$ as a function of strain would probably not be observed if the samples were slipping on the substrate with temperature cycling. Furthermore, the contacts placed on top the samples (Fig. 1 b)) would crack along the edge if slippage was occurring. Instead, the samples appear to commonly survive with contacts intact over multiple temperature cycles. (Data from the ones which did not survive were not used in the analysis). However, additional strain may arise from the contacts themselves, which are not taken into account, and their effect is hard to estimate. Nevertheless the systematic behavior of $T_{c2}$ with strain suggests that the contribution of the contacts is also systematic, possibly leading to a systematic error in the absolute value of applied stress.

Comparison of the present experiments of in-plane strain with results of hydrostatic pressure [4] leads to some interesting conclusions. In Figure 3 b) the top axis shows the estimated hydrostatic pressure $P$, in MPa, calculated using the relation $P = v^2 \rho e$, where $\rho$ is the density (6.86 g/cm$^3$), $v$ is the sound velocity (1850 m/s) [9]. Although the hydrostatic pressure experiments by Sipos et al. [4] lead to much larger strains, our in-plane compressive strain appears to systematically *increase* $T_{c1}$ (Fig. 3), while the pressure experiments show the opposite trend. One obvious conclusion would be that the in-plane strain is not responsible for the decrease in $T_{c2}$, but rather, it is the *c*-axis strain which plays a role in altering the properties described by Sipos et al. [4], including pressure-induced superconductivity. This can be verified experimentally by applying *c*-axis stress.

Turning to $T_H$, this shows an opposite trend to $T_{c2}$, decreasing with increasing compressive strain. This indicates that strain influences relaxation of the *H* state by a different mechanism than the *NC-C* transition at $T_{c2}$. The *H* state has been proposed to be a CDW state which is possibly related to the *NC* state above $T_{c2}$ [1] but the details of the charge ordering have not yet been experimentally determined. Yet the present results suggest that the two states are quite different from the point of view of strain.

# Conclusions

The presented measurements give two important insights. The first pertains to the strong and systematic increase of $T_H$ on tensile strain, which is not displayed by $T_{c2}$, highlighting the difference between the two transitions. While both transitions are thought to be of first order, it appears that the *H-C* state transition has a different character than the *NC-C* state transition. Increasing tensile strain appears to rapidly increase $T_H$, suggesting a method for increasing the stability of the *H* state at high temperatures by choice of appropriate substrate. The second insight is that $T_{c2}$ displays the opposite dependence of $T_{c2}$ on compressive in-plane strain to previously reported hydrostatic pressure measurements. The implication of the latter is that *c*-axis strain plays an important role in determining the functional properties, and particularly pressure-induced superconductivity of this remarkable system.

Figure 1.

a) A schematic picture of the experimental setup and a typical sample on sapphire substrate used in this investigation. The laser pulse is incident through the substrate.
b) A photograph of a typical sample (100 nm thick $1T$-$TaS_2$ on sapphire).

Figure 2.

a) The temperature dependence of the lattice constants of selected substrates and $1T$-$TaS_2$. The lines are fitted data.
b) The differential strain in-plane applied by different substrates on $1T$-$TaS_2$. Sapphire, MgO and quartz apply tensile strain upon cooling, while $CaF_2$ show compressive strain. BK7 causes a small compressive strain on cooling, and a small tensile strain on warming. Fitted curves (from a) are shown for clarity.

Figure 3.

a) The temperature dependence of the resistivity of $1T$-$TaS_2$ on different substrates near *NC-C* state transition on cooling and warming.
b) The effect of strain *e* on the *NC-C* state transition temperature on cooling ($T_{c2}^c$) and warming ($T_{c2}^w$) measured by resistivity (normalized to the room temperature value). The strain is calculated at the actual transition temperature in each case (from Fig. 2). The top scale refers to the equivalent hydrostatic pressure in MPa. The error bars indicate standard deviation.

Figure 4.

a) The temperature dependence of the resistivity of $1T$-$TaS_2$ on different substrates in the *H* state on warming.
b) The transition temperature $T_H$ as a function of strain *e*. The strain is calculated at the actual transition temperature (from Fig. 2). The top scale refers to the equivalent hydrostatic pressure in MPa. The error bar indicates the standard deviation.

Figure 1

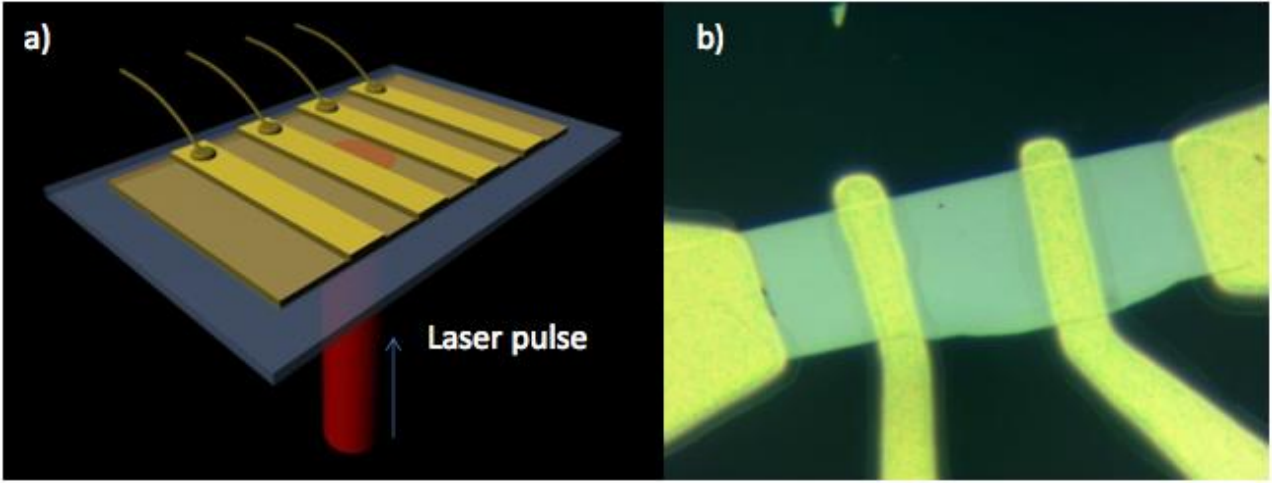

Figure 2

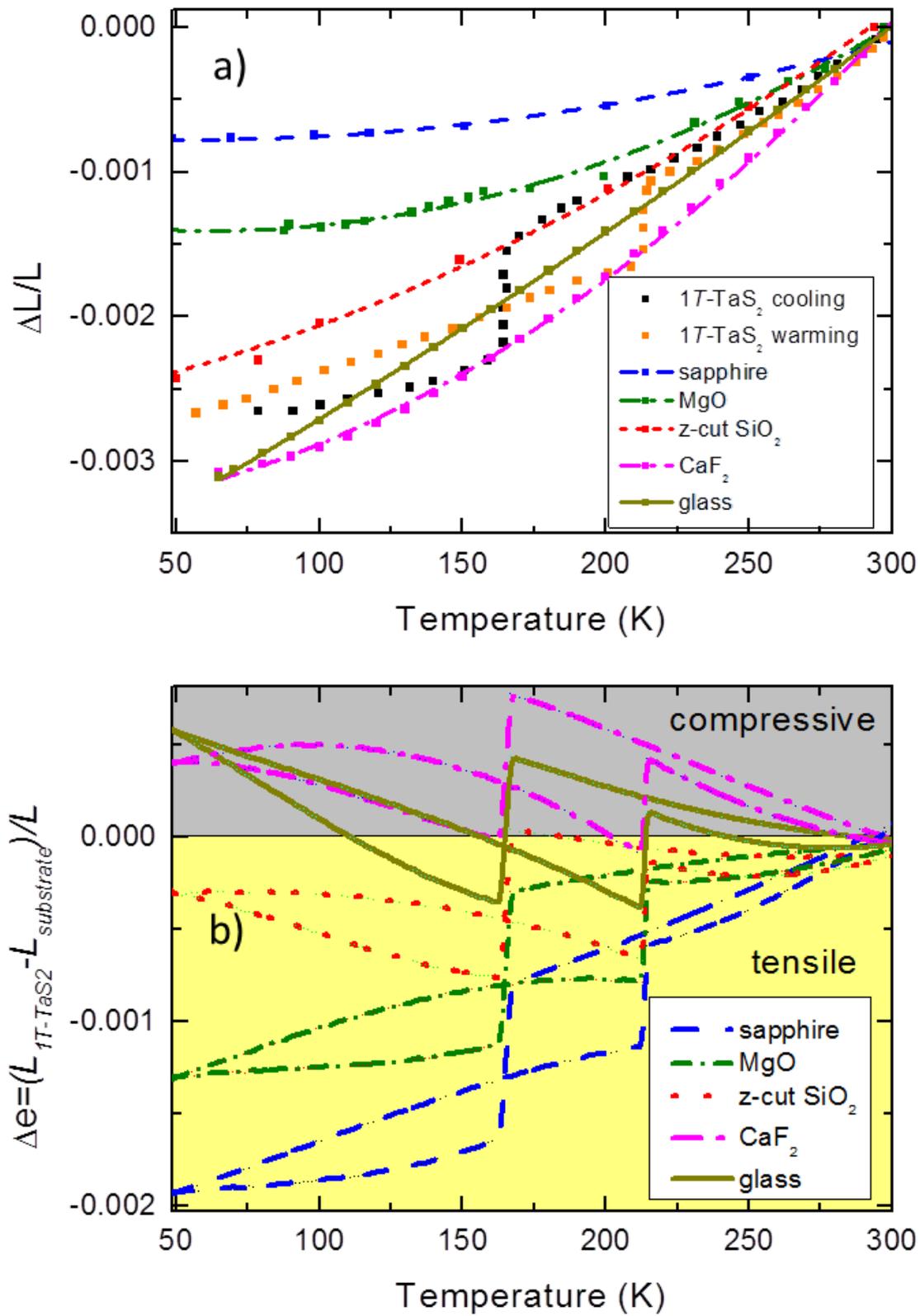

Figure 3

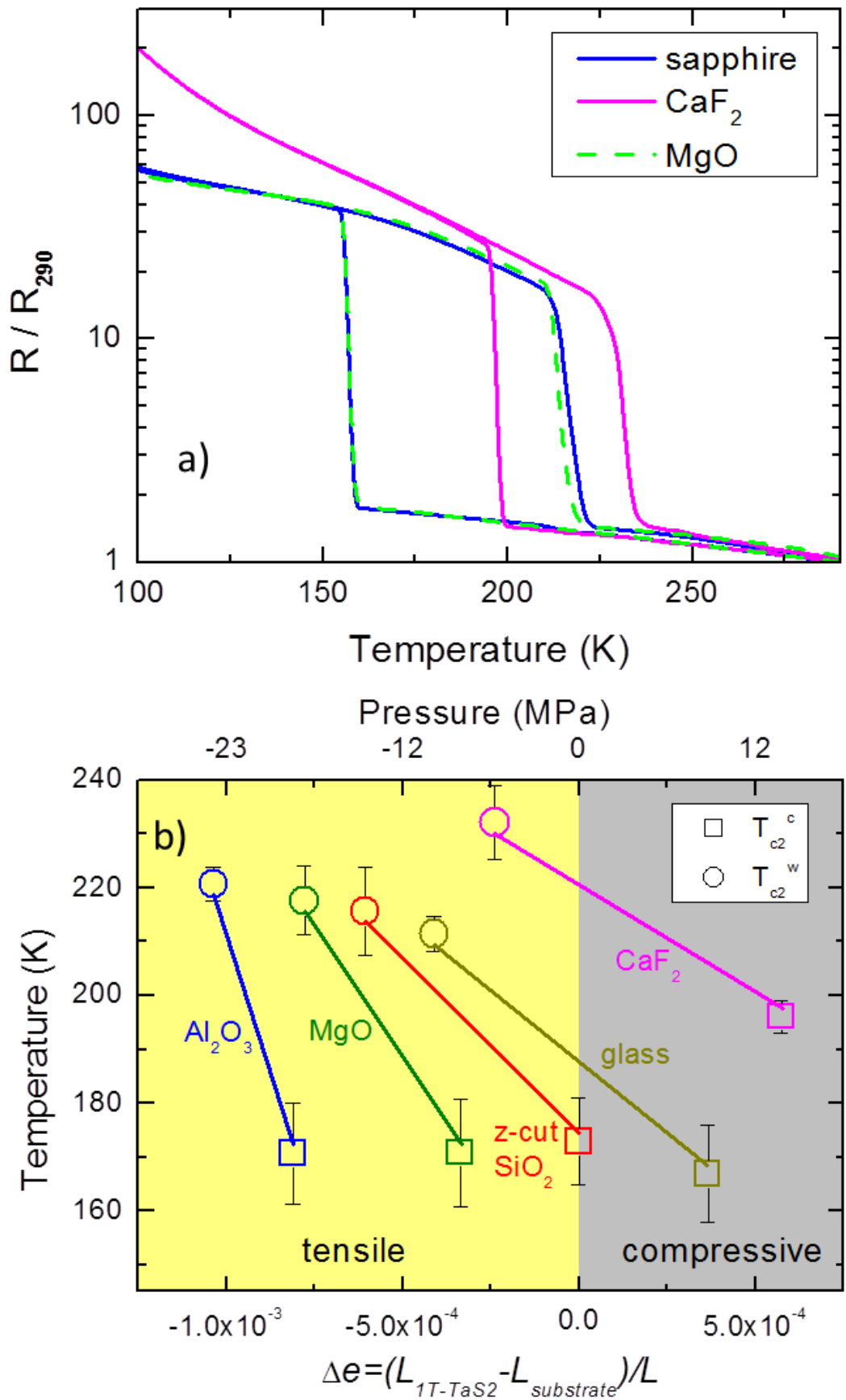

Figure 4

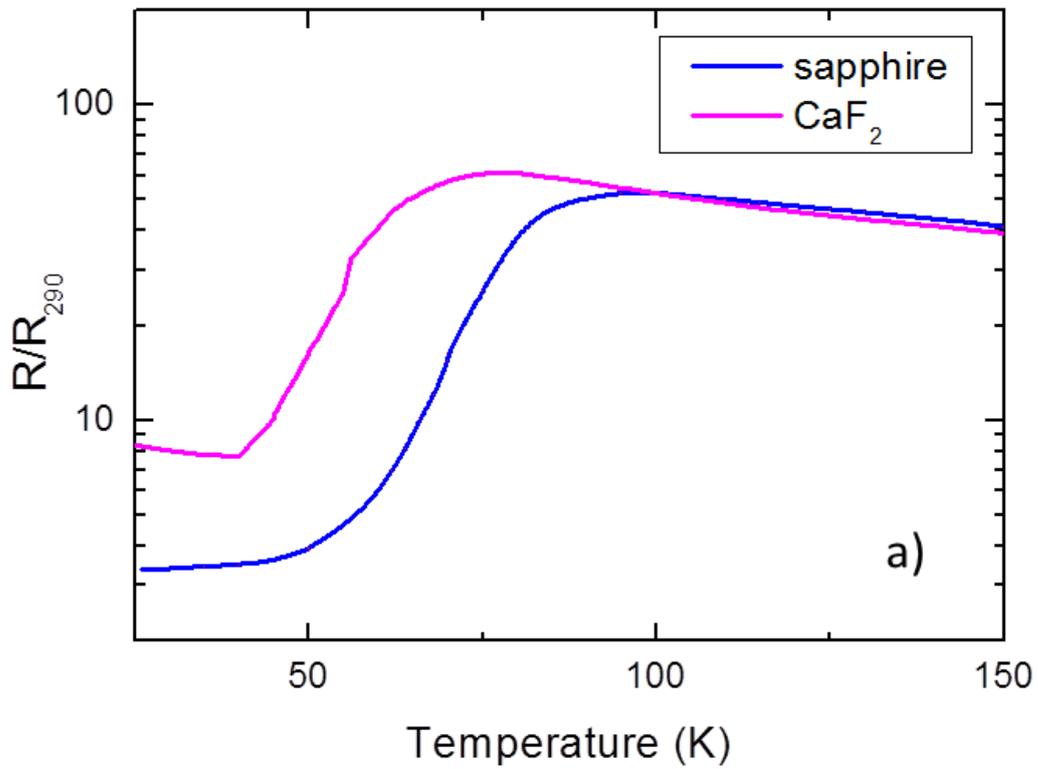

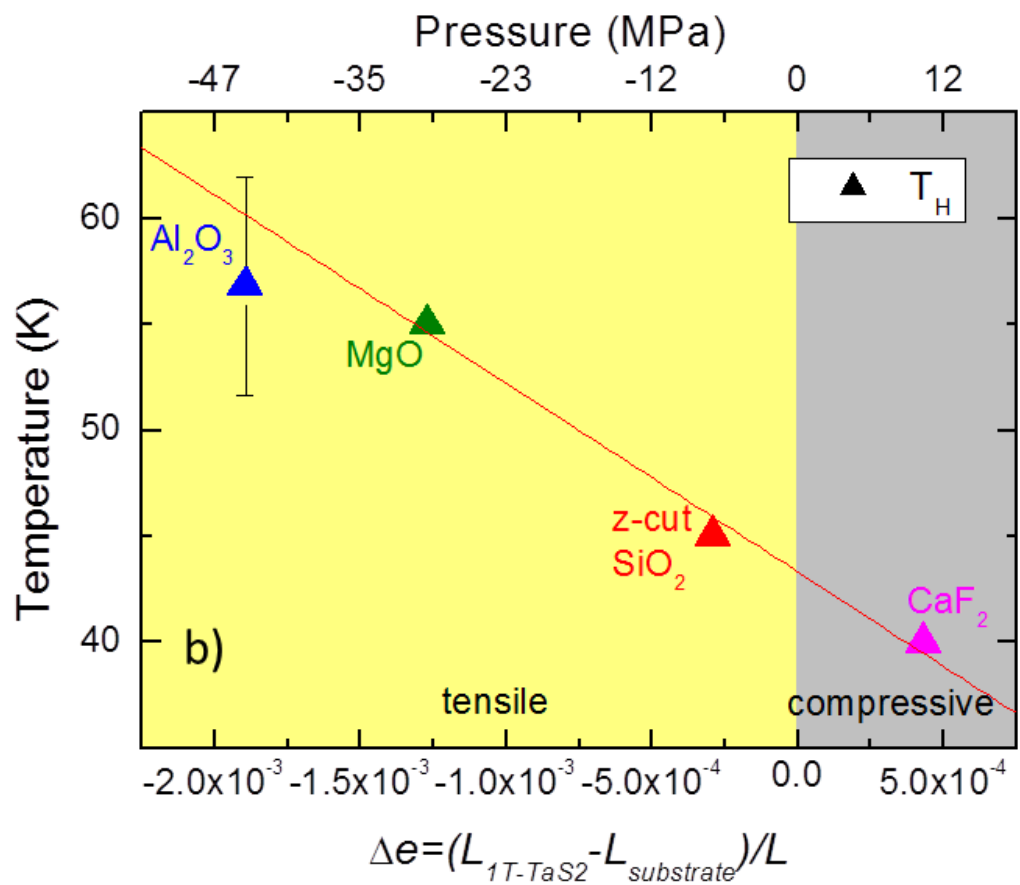